\documentclass[final,5p,times,twocolumn,authoryear]{elsarticle}  
\usepackage{colortbl} 
\usepackage[normalem]{ulem}
\usepackage{amssymb}
\usepackage{graphicx}
\usepackage{txfonts}
\usepackage{array}
\journal{New Astronomy}
\usepackage[figuresright]{rotating}
\begin{document}
\def\astrobj#1{#1}
\def\teff{$T_{\rm eff}$}
\def\logg{$\log g$}
\def\micro{$\xi$ }
\def\kms{km s$^{-1}$}
\def\p{$\pm$}
\def\vsini{$v\sin i$}
\begin{frontmatter}
  \title{Revisiting the orbital motion of WR\,138}
\author[gr]{Gregor Rauw}
\ead{g.rauw@uliege.be}
\author[gr]{Ya\"el Naz\'e\fnref{fn1}}
\fntext[fn1]{Senior Research Associate FRS-FNRS (Belgium).}
\author[gr]{Eric Gosset\fnref{fn2}}
\fntext[fn2]{Honorary Research Director FRS-FNRS (Belgium).}
 \address[gr]{Space sciences, Technologies and Astrophysics Research (STAR) Institute, Universit\'e de Li\`ege, All\'ee du 6 Ao\^ut, 19c, B\^at B5c, 4000 Li\`ege, Belgium}
 \begin{abstract}
The optical spectrum of WR~138 exhibits emission lines typical of a WN6o star and absorption lines from a rapidly-rotating OB star. Using a large set of spectroscopic data, we establish a new orbital solution of the WN6o star based on the radial velocities of highly-ionized nitrogen lines. We show that the WN6o star moves on a 4.3\,yr orbit with a comparatively low eccentricity of 0.16. The radial velocities of the OB star display considerable scatter. Our best estimates of the velocities of He\,{\sc i} absorption lines result in a mass-ratio of $m_{\rm WN6o}/m_{\rm OB} = 0.53 \pm 0.09$. We disentangle the spectra of the two stars and derive a projected rotational velocity of $v\,\sin{i} = 350 \pm 10$\,km\,s$^{-1}$ for the OB star. Our best orbital parameters, combined with the Gaia parallax of WR~138, are at odds with a previous interferometric detection of the companion, suggesting that there is either a bias in this detection or that WR~138 is actually a triple system. 
  \end{abstract}
\end{frontmatter}
\section{Introduction\label{intro}}
There is consensus that classical Wolf-Rayet (WR) stars are evolved descendants of massive OB-type stars that have lost their outer hydrogen-rich layers \citep{Cro07}. How exactly this evolution proceeds is not clear though: both wind mass-loss and binary interaction could lead to the formation of WR stars. The incidence of known binaries among Galactic WR stars led to the suggestion that $\geq 70$\% of their progenitor O-type stars are in binary systems that are close enough to enable a Roche lobe overflow (RLOF) interaction over their lifetime \citep{Van98}. Over the last two decades, intensive studies of O stars indeed revealed that the majority of them are part of binary or higher multiplicity systems \citep[][and references therein]{San12}. The mass and angular momentum transfers that occur during a RLOF interaction can result in a significant spin-up of the mass gainer \citep{Pac81}. In WR + O binaries that have evolved through binary interaction, the O star should thus be a rapid rotator. This expectation was confirmed observationally for a sample of eleven WR + O systems \citep{Sha17}, though the rotation rates were found to be significantly subcritical \citep{Van18}. Probably one of the most extreme cases to investigate in this context is \astrobj{WR~138} (= HD~193\,077), the target of the present study.

Although the presence of absorption lines in the spectrum of WR~138 was recognized a long time ago \citep{Hil45}, the multiplicity of this star has been an open issue for many years. \citet{Massey} found no evidence for an orbital motion of the emission lines with an amplitude larger than 30\,km\,s$^{-1}$ and for periods of less than six months. He accordingly suggested that the broad absorption lines, for which he estimated $v\,\sin{i} \simeq 500$\,km\,s$^{-1}$, would be intrinsic to the WN star. Using a more extensive series of photographic spectra, \citet{Lamontagne} noted low-amplitude radial velocity (RV) variations of the N\,{\sc iv} $\lambda$\,4058 emission line on periods of 2.3238\,days and either of 1763 or 1533\,days. Yet, the short period variations were seen neither in the He\,{\sc ii} $\lambda$\,4686 emission line nor in the absorption lines. Instead, the RVs of the He\,{\sc ii} $\lambda$\,4686 emission were seen to vary on the same long period as those of the N\,{\sc iv} $\lambda$\,4058 line, and the absorption line RVs were found to vary in antiphase also on that period. \citet{Lamontagne} therefore classified WR~138 as a triple system consisting of a WN6 star orbited every 2.32\,days by an unseen companion, which they tentatively proposed to be a neutron star, and an O-type star on a much wider orbit.
The existence of the 2.32\,day cycle was not confirmed in subsequent RV studies \citep{Ann90,Pal13}. Moreover, WR~138 does not display the strong and hard X-ray emission one would expect from a neutron star in close orbit to a WN star. Instead, the X-ray emission was found to be typical of wide WR + OB colliding wind binaries \citep{Pal13}. The most recent orbital solution of the WN star was published by \citet{Ann90}, who inferred an orbital period of $1538 \pm 14$\,days, a RV amplitude of $K \sim 30$\,km\,s$^{-1}$ and an eccentricity of $e \sim 0.3$. Yet, different emission lines led to different values of $K$ and $e$, and \citet{Ann90} could not measure the RVs of the absorption lines.
Further information regarding the multiplicity of WR~138 was gathered via $H$-band {\it CHARA} interferometry that allowed to resolve the star into two objects separated by 12.4\,mas \citep{Rich16}. \citet{Rich16} thus suggested that WR~138 might have evolved through a previous mass-transfer episode that created the WR star and led to a spin-up of the O-type star.

In the present work, we revisit the orbit of WR~138, using a large set of optical spectra collected over twelve years. The data used in our study are described in Sect.\,\ref{obs}. The analysis of the spectroscopic and photometric data are presented in Sects.\,\ref{spectro} and \ref{photom} respectively. The implications of our results are finally discussed in Sect.\,\ref{disc}, and our conclusions are given in Sect.\,\ref{conclusion}.  

\section{Observations \label{obs}}
Between 2011 and 2022, we observed WR~138 with the Aur\'elie spectrograph \citep{Gillet} at the 1.52\,m telescope of the Observatoire de Haute Provence (OHP) in France. Data were collected once per year (except in 2014, 2015 and 2021) in the framework of observing runs lasting for six nights each. Different gratings and different wavelength domains were used. The data from 2011 and 2012 were previously used in the work of \citet{Pal13}. The most frequently covered wavelength domain extends from 4448 to 4886\,\AA\ and was observed at a resolving power of $\simeq 10000$. Until 2018, the detector was an EEV CCD with $2048 \times 1024$ pixels. From 2019 on, it was replaced by an Andor CCD camera with $2048 \times 512$ pixels. Both CCDs had pixel sizes of 13.5\,$\mu$m squared. Typical integration times of individual exposures were 30\,min. The Aur\'elie spectra were reduced using version 17FEBpl\,1.2 of the {\sc midas} software developed at ESO.

Since 2016, we monitored WR~138 more regularly with the fully robotic 1.2~m TIGRE telescope \citep{Sch14,Gon22} located at La Luz Observatory near Guanajuato (Mexico). TIGRE collects echelle spectra with the refurbished HEROS spectrograph \citep{Kau98,Sch14,Gon22} which provides a spectral resolving power of 20\,000 over the optical range from 3760 -- 8700\,\AA, except for a small gap between 5660 and 5770\,\AA. Individual integration times were typically 1\,hour. The data were reduced with the HEROS reduction pipeline \citep{Mit11,Sch14}.

%In July 2016, we further collected one observation of WR~138 with the echelle spectrograph on the 2.12\,m telescope at Observatorio Astron\'omico Nacional of San Pedro M\'artir (SPM) in Mexico. These data cover the spectral region from 3900 to 7200\,\AA\ at a resolving power of $\simeq 18000$. The integration time was 45\,min. The data reduction was done with {\sc midas}.

High-precision space-borne photometry of WR~138 was collected with the Transiting Exoplanet Survey Satellite \citep[{\it TESS},][]{TESS} during sectors 14 (18 July - 15 August 2019), 15 (15 August - 11 September 2019), 41 (23 July - 20 August 2021), and 55 (5 August - 1 September 2022). All observations were taken with {\it TESS} camera 1, except for the last sector, where WR~138 was observed with camera 3. For sectors 14 and 15 the data were obtained at a cadence of one observation every 30\,min, whilst it was one observation every 10\,min for sectors 41 and 55. We retrieved the corresponding full frame images from the Mikulski Archive for Space Telescopes (MAST) portal\footnote{http://mast.stsci.edu/}, and extracted aperture photometry light curves with the {\tt Lightkurve}\footnote{https://docs.lightkurve.org} software. The background correction was done using either the median flux of the background pixels (defined as pixels with flux values below the median of a $50 \times 50$ pixels cut-out) or performing a principal component analysis (PCA) with five components. Both methods yield very similar results. In the following we focus on the light curve obtained via the PCA background correction. The stellar fluxes corrected for the background were finally converted into magnitudes and the mean magnitude of the corresponding sector was subtracted. 

\section{Radial velocities \label{spectro}}
The optical spectrum of WR~138 features many strong and broad emission lines, typical of a WN6o star, where the 'o' tag indicates the absence of H in the spectrum \citep{Smi96,Cro11}. Previously, WR~138 was classified as WN5o + B? by \citet{Smi96}, WN6 + O9-9.5\,I-II by \citet{Ann90}, WN6o by \citet{Cro11}, and WN5o + O9\,V by \citet{Rich16}. The relative strengths of the N\,{\sc iii}, N\,{\sc iv} and N\,{\sc v} lines clearly position WR~138 at the boundary between the WN5 and WN6 spectral types. Hereafter, we will adopt the WN6o classification for the Wolf-Rayet star, following the revised classification criteria of \citet{Cro11}. In addition to these emission lines, the spectrum displays broad He\,{\sc i} absorption lines that are not P-Cygni absorption troughs and are attributed to the OB star.

\subsection{The orbit of the WN6o star}
The majority of the emission lines are either blends of different ions (e.g.\ the N\,{\sc v} $\lambda$\,4620 line that is blended with the N\,{\sc iii} $\lambda\lambda$\,4634 - 42 complex) or occasionally display a complex morphology of their peak (e.g.\ He\,{\sc ii} $\lambda\lambda$\,4542, 4686, 5412) thus making them less suitable for RV measurements. Indeed, as for many Wolf-Rayet stars, the broad emission lines of WR~138 display variable subpeaks \citep{Lep99,Pal13} that could actually affect the determination of the RV of the WN star. Ideally, we aim at measuring the RVs of lines from ions with a high ionization potential as they should arise deeper within the wind \citep{Her00}, thereby better reflecting the true RV variations of the WN6o star. Moreover, these N\,{\sc iv} and N\,{\sc v} lines were empirically found to display the smallest scatter in high-cadence observations of WR~138 \citep{Dsilva}, suggesting that they are less affected by the wind inhomogeneities that strongly impact the profiles of He\,{\sc ii} emission lines \citep{Lep99}. In some WR + OB binaries, additional emission can arise from the radiatively cooling material in the post-shock region of the colliding wind interaction zone. Such an extra emission component could distort the line profile and bias the RV determination. The X-ray spectrum of WR~138 indeed suggests the presence of a wind interaction \citep{Pal13}. Considering the wind parameters inferred by \cite{Rich16} and the orbital parameters from our orbital solution below, we can estimate a cooling parameter $\chi = 3.0/\sin{i}$ for the shocked wind of the WN star at periastron (i.e., at the orbital phase when cooling should be most efficient). This $\chi$ parameter provides the ratio between the cooling time and the escape time from the shock region \citep{SBP}. A value larger than 1 implies an adiabatic interaction zone, where radiative cooling does not play a significant role. We thus expect very little extra optical line emission from the wind interaction zone in WR~138.  

\begin{table}
  \caption{RVs of WR~138 as measured on our data. In the last column, 'A' stands for the mean RVs obtained during a six-night OHP 1.52\,m + Aur\'elie observing run, 'T' indicates spectra taken with TIGRE + HEROS.  \label{TabRVs}}
  \begin{tabular}{c r r c}
    \hline
    HJD-2450000 & \multicolumn{1}{c}{RV$_{\rm WN}$} & \multicolumn{1}{c}{RV$_{\rm OB}$} & Instrument\\
    \hline
5827.801 & $ -30.5 \pm 5.9 $ & $43.8 \pm 13.6$ & A \\
6093.979 & $ 33.1 \pm 7.1 $ & $-27.3 \pm 8.8$ & A \\
6457.929 & $ 37.4 \pm 6.5 $ & $-32.6 \pm 9.2$ & A \\
7508.938 & $ -9.2 \pm 15.7 $ & $-5.0 \pm 6.8$ & T\\
7550.972 & $ 19.1 \pm 3.1 $ & $-15.1 \pm 9.3$ & A \\
7842.971 & $ 26.9 \pm 6.5 $ & $-18.1 \pm 2.6$ & T\\
7890.918 & $ 22.4 \pm 10.1 $ & $-10.0 \pm 18.1$ & T\\
7927.850 & $ 29.8 \pm 8.2 $ & $-33.5 \pm 26.0$ & T\\
8004.808 & $ 26.0 \pm 5.1 $ & $-37.8 \pm 6.1$ & A \\
8016.667 & $ 20.3 \pm 5.1 $ & $-7.8 \pm 16.6$ & T\\
8053.570 & $ 17.7 \pm 2.2 $ & $3.2 \pm 29.7$ & T\\
8209.934 & $ 4.7 \pm 11.9 $ & $-6.6 \pm 8.1$ & T\\
8240.945 & $ -2.3 \pm 7.9 $ & $-0.7 \pm 6.2$ & T\\
8262.938 & $ 2.4 \pm 5.9 $ & $-14.9 \pm 14.6$ & T\\
8303.918 & $ -0.7 \pm 6.4 $ & $-3.4 \pm 13.9$ & T\\
8354.926 & $ -13.5 \pm 4.5 $ & $-13.7 \pm 7.3$ & A \\
8574.933 & $ -32.4 \pm 4.9 $ & $51.0 \pm 8.9$ & T\\
8604.943 & $ -44.5 \pm 11.9 $ & $12.1 \pm 13.1$ & T\\
8644.856 & $ -40.4 \pm 2.3 $ & $23.1 \pm 17.6$ & T\\
8648.834 & $ -43.1 \pm 8.4 $ & $16.8 \pm 14.3$ & T\\
8675.917 & $ -48.8 \pm 8.4 $ & $12.8 \pm 28.3$ & T\\
8705.800 & $ -47.3 \pm 10.8 $ & $5.3 \pm 1.9$ & T\\
8719.840 & $ -40.1 \pm 10.2 $ & $15.9 \pm 5.9$ & A \\
8738.705 & $ -45.5 \pm 9.1 $ & $15.7 \pm 15.8$ & T\\
8767.600 & $ -42.8 \pm 5.4 $ & $14.9 \pm 22.9$ & T\\
8810.572 & $ -37.9 \pm 3.9 $ & $17.1 \pm 19.9$ & T\\
8941.952 & $ -16.7 \pm 7.4 $ & $19.2 \pm 5.7$ & T\\
8980.924 & $ -16.6 \pm 7.6 $ & $20.7 \pm 7.7$ & T\\
9035.846 & $ 5.2 \pm 7.2 $ & $-16.1 \pm 5.4$ & T\\
9096.661 & $ 16.0 \pm 10.6 $ & $-12.3 \pm 12.9$ & T\\
9127.640 & $ 5.9 \pm 8.5 $ & $-9.7 \pm 12.6$ & T\\
9143.590 & $ 39.0 \pm 14.9 $ & $-9.3 \pm 15.0$ & T\\
9174.573 & $ 20.7 \pm 13.6 $ & $16.4 \pm 3.5$ & T\\
9307.949 & $ 36.3 \pm 8.0 $ & $-38.7 \pm 13.1$ & T\\
9443.683 & $ 33.7 \pm 14.5 $ & $-5.1 \pm 5.8$ & T\\
9499.587 & $ 36.9 \pm 4.7 $ & $-0.5 \pm 6.6$ & T\\
9524.594 & $ 29.0 \pm 5.3 $ & $-10.0 \pm 8.2$ & T\\
9670.956 & $ 22.6 \pm 9.6 $ & $-16.6 \pm 7.5$ & T\\
9686.893 & $ 15.9 \pm 9.9 $ & $0.3 \pm 12.6$ & T\\
9687.887 & $ 25.9 \pm 7.8 $ & $8.2 \pm 12.6$ & T\\
9689.957 & $ 36.1 \pm 3.4 $ & $13.3 \pm 12.6$ & T\\
9691.963 & $ 18.9 \pm 8.0 $ & $-9.7 \pm 12.6$ & T\\
9693.887 & $ 32.3 \pm 2.6 $ & $-11.6 \pm 12.6$ & T\\
9853.787 & $ -8.7 \pm 5.5 $ & $-5.2 \pm 4.2$ & A \\
10043.958 & $-22.5 \pm 11.7$ & $5.9 \pm 3.0$ & T \\
\hline
  \end{tabular}
\end{table}

\begin{figure}%[thb]
  \begin{center}
    \resizebox{8.5cm}{!}{\includegraphics{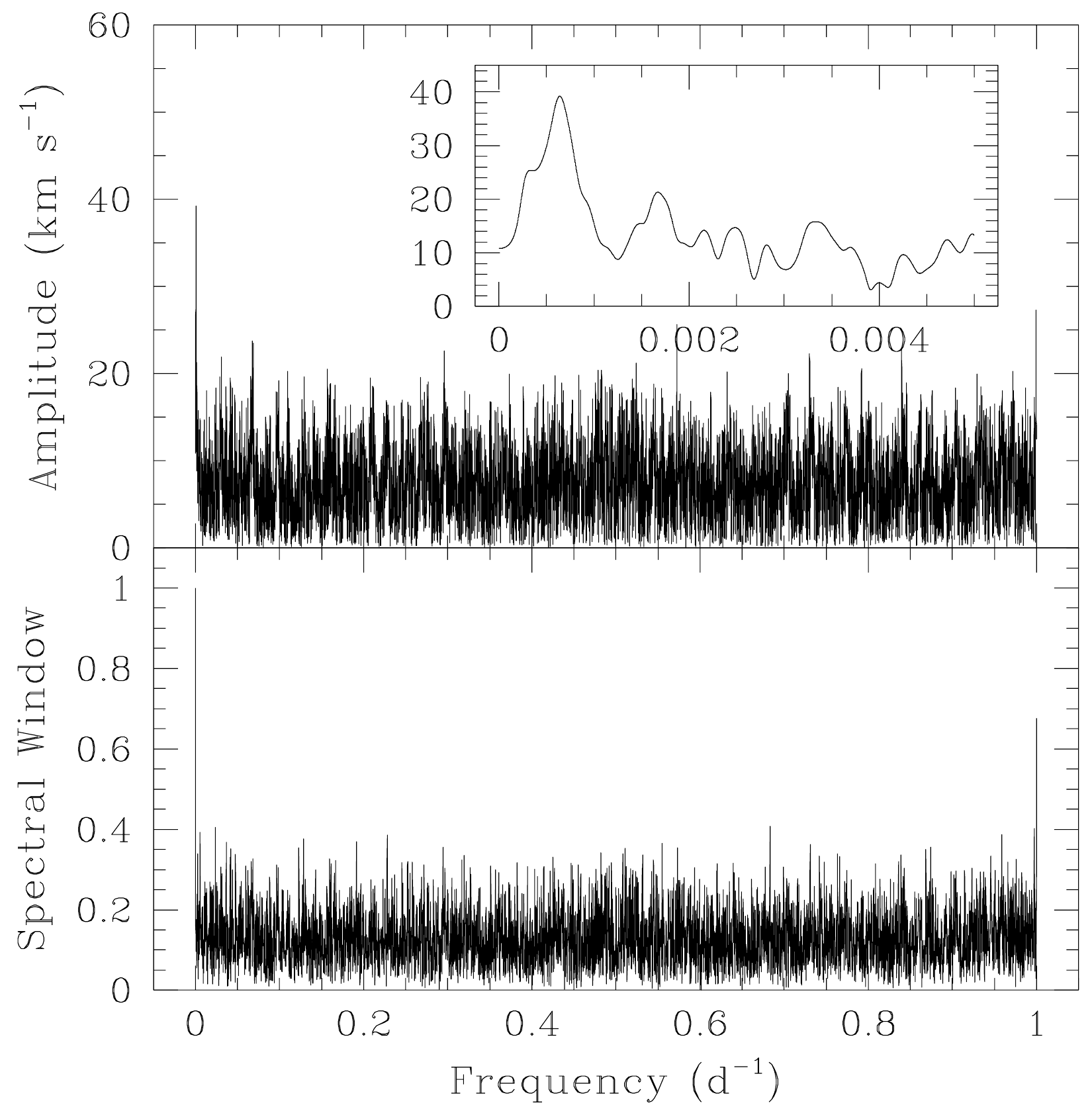}}
  \end{center}
  \caption{Fourier spectrum of the RV$_{\rm WN}$ time series (top) and  spectral window of the data (bottom). The insert in the top frame provides a zoom on the low frequency part of the spectrum.\label{specFour}}
\end{figure}

To determine the RVs of the WN6o star, we thus selected several relatively isolated lines of highly-ionized nitrogen: N\,{\sc iv} $\lambda\lambda$\,4058, 5204, 7103 - 7129, and N\,{\sc v} $\lambda\lambda$\,4604, 4945. The N\,{\sc iv} $\lambda\lambda$\,7103 - 7129 feature is a complex multiplet but shows a well-defined peak near $\lambda$\,7117, thus making it suitable for RV measurements. The N\,{\sc v} $\lambda$\,4604 line is part of a doublet together with N\,{\sc v} $\lambda$\,4620. However, since the latter component is severely blended with the N\,{\sc iii} $\lambda\lambda$\,4634 - 42 complex, only the N\,{\sc v} $\lambda$\,4604 could be measured. We determined the line centroids of these five emissions on the TIGRE data using Gaussian fits to the entire line profile. For each line, we then subtracted the mean RV over the full TIGRE campaign. Finally, for each TIGRE observation, we took the mean of these corrected RVs of the nitrogen lines. In view of the nearly even sampling of the orbital cycle by the TIGRE data and assuming a nearly circular orbit, the ensuing orbital solution should have a systemic velocity close to 0\,km\,s$^{-1}$ by construction. The dispersion about the resulting RV$_{\rm WN}$ was used as an indicator of the uncertainty\footnote{Throughout this paper, quoted errors correspond to $1\,\sigma$ intervals.}. In the OHP spectra, only the N\,{\sc v} $\lambda$\,4604 line is covered. We subtracted the mean RV of this line, as determined on the TIGRE data, from the OHP measurements. We then averaged all the corrected RVs obtained during the same six-night OHP observing run. The resulting RV$_{\rm WN}$ values are listed in Table\,\ref{TabRVs}.

To establish the orbital period, we applied the modified Fourier periodogram algorithm of \citet{Hec85} and \citet{Gos01} to the corresponding time series of RV$_{\rm WN}$ (see Fig.\,\ref{specFour}). This Fourier method explicitly accounts for the highly uneven sampling of the time series. The time series spans an interval of 4216\,days, leading to a natural width of the peaks in the periodogram of $2.37\,10^{-4}$\,d$^{-1}$. The highest peak in the periodogram is found at a frequency of $(6.40 \pm 0.12)\,10^{-4}$\,d$^{-1}$, corresponding to an orbital period of $(1563 \pm 29)$\,days, i.e.\ 4.3\,yrs. The quoted uncertainty is obtained assuming that the position of the highest peak in the periodogram can be determined with an accuracy of 1/20 of the natural width of the peak.  

\begin{figure}%[thb]
  \begin{center}
    \resizebox{8.5cm}{!}{\includegraphics{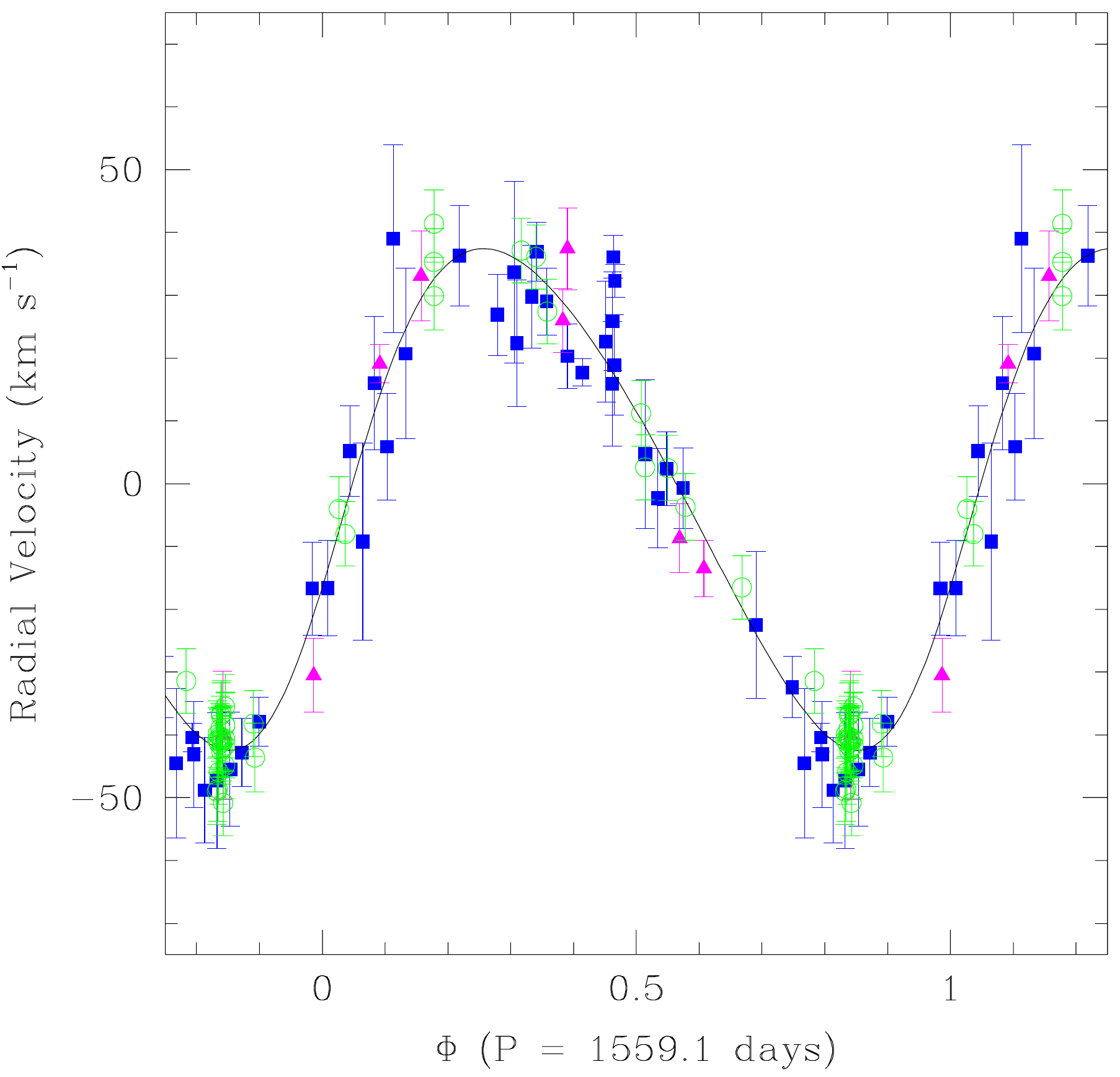}}
  \end{center}
  \caption{SB1 orbital solution of the WN star as determined from the RVs of the highly-ionized nitrogen lines RV$_{\rm WN}$ combined with the data of \citet{Dsilva}. Blue squares stand for the TIGRE data, while the magenta triangles indicate OHP data and green circles are RVs from \citet{Dsilva}.\label{solobsSB1}}
\end{figure}
We used the Li\`ege Orbital Solution Package \citep[LOSP,][]{San06} to establish an SB1 orbital solution for the WN star (Fig.\,\ref{solobsSB1}), taking the above determination of the orbital period as a first guess, but allowing LOSP to adjust its value. All data points with a nominal uncertainty below 10\,km\,s$^{-1}$ were assigned a weight of 1.0, whilst those with larger uncertainties were given a weight of 0.5. The resulting orbital elements are quoted in Table\,\ref{orbel}.

We also used the 40 RV measurements of the WN6o star published by \citet{Dsilva} for what they call weak N\,{\sc v} lines (which includes N\,{\sc v} $\lambda\lambda$\,4604, 4620, 4945). Those RVs are relative values as they were obtained via a cross-correlation with a template spectrum built from their highest quality spectra. With respect to our RV$_{\rm WN}$ values, we found a shift in systemic velocity of 45.9\,km\,s$^{-1}$, that we then applied to compute a combined orbital solution (see Fig.\,\ref{solobsSB1}). Overall, we find an excellent agreement between our RVs and those of \citet{Dsilva}. Compared to the solutions of \citet{Ann90}, our orbital solution yields a $\sim 25$\% larger amplitude (39.9\,km\,s$^{-1}$ versus 30.6 -- 34.3\,km\,s$^{-1}$) and a significantly lower eccentricity ($0.16$ versus 0.29 -- 0.35). The rather large value of the mass function clearly points towards a companion that must be a relatively massive object. 

\begin{table}
  \caption{SB1 orbital elements of the WN6o star in WR~138. The $\gamma$ velocity is consistent with 0 by construction of the RVs. It does not reflect the true apparent systemic velocity of the WN6o star. $t_0$ stands for the time of periastron passage, while $\omega$ is the argument of periastron. $f(m)$ is the mass function.\label{orbel}}
  \begin{center}
  \begin{tabular}{c c c}
    \hline
    & RV$_{\rm WN}$ & RV$_{\rm WN}$ + RV$_{\rm Dsilva}$  \\ 
    \hline
   $P_{\rm orb}$ (days) & $1553 \pm 14$ & $1559 \pm 10$  \\  
   $\gamma$ (km\,s$^{-1}$) & $-1.1 \pm 1.2$ & $-0.3 \pm 0.7$ \\
   $e$ & $0.15 \pm 0.04$ & $0.16 \pm 0.03$ \\
   $K$ (km\,s$^{-1}$) & $40.6 \pm 1.7$ & $39.9 \pm 1.0$ \\
   $\omega$ ($^{\circ}$) & $233 \pm 16$ & $250 \pm 9$ \\
   $a_{\rm WN6o}\,\sin{i}$ (R$_{\odot}$) & $1233 \pm 55$ & $1215 \pm 32$\\
   $t_0$ (HJD-2450000) & $7343 \pm 66$ & $7408 \pm 39$ \\
   $f(m)$ (M$_{\odot}$) & $10.4 \pm 3.8$ & $9.9 \pm 2.6$ \\
   rms (km\,s$^{-1}$) & 7.5 & 6.0 \\
   \hline
  \end{tabular}
  \end{center}
\end{table}

\begin{figure}%[thb]
  \begin{center}
    \resizebox{8.5cm}{!}{\includegraphics{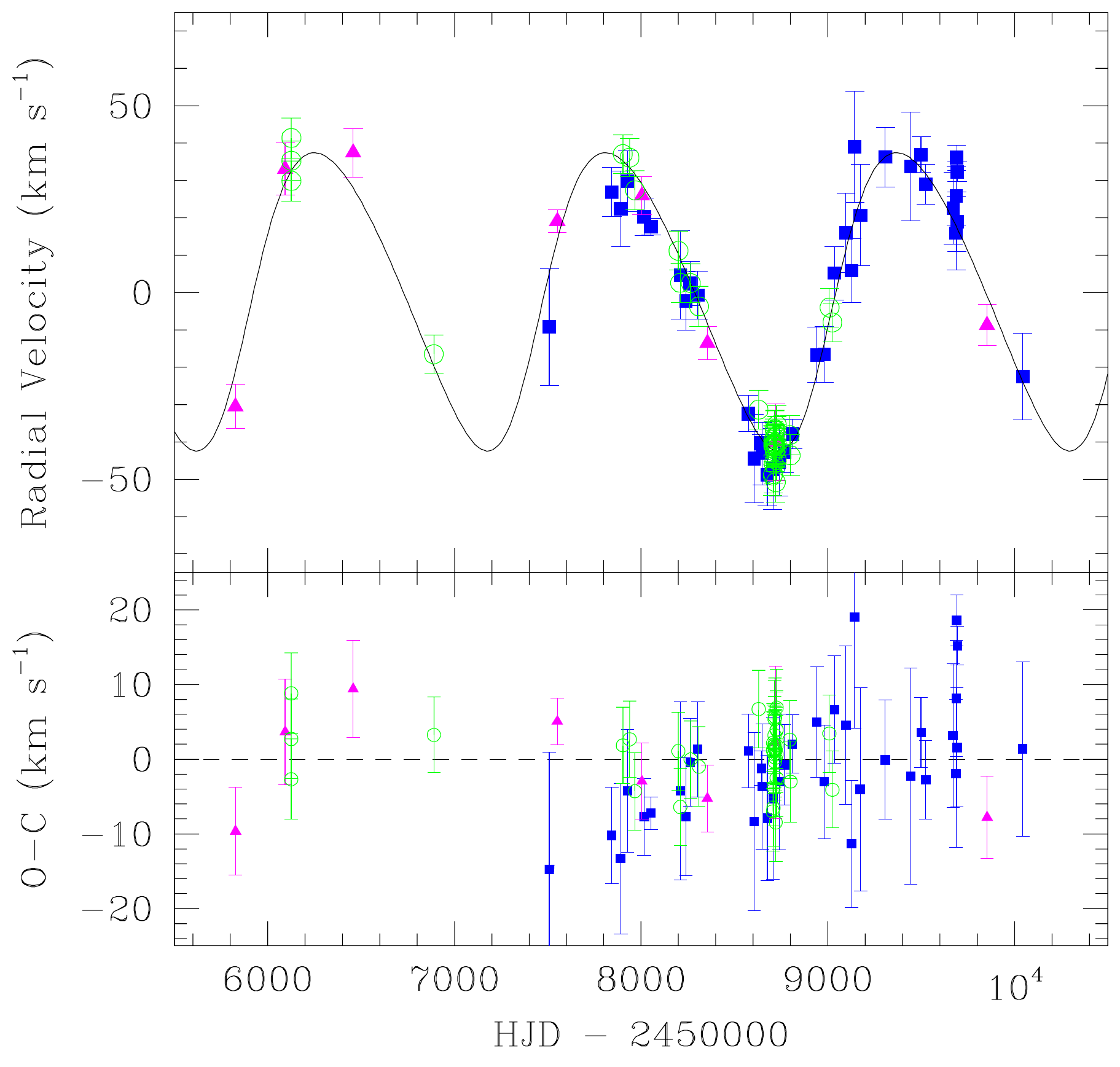}}
  \end{center}
  \caption{RV$_{\rm WN}$ combined with the data of \citet{Dsilva} (top panel) and associated O-C residuals with respect to our orbital solution (bottom panel) as a function of time. The symbols have the same meaning as in Fig.\,\ref{solobsSB1}. \label{OC}}
\end{figure}

Our complete dataset covers a bit less than three full orbital cycles of WR 138. To look for possible cycle to cycle variations, we inspected both the RVs and the O-C residuals of our orbital solution as a function of time (Fig.\,\ref{OC}). No significant systematic trends were found.

\begin{figure}[h!]
  \begin{center}
    \resizebox{8.5cm}{!}{\includegraphics{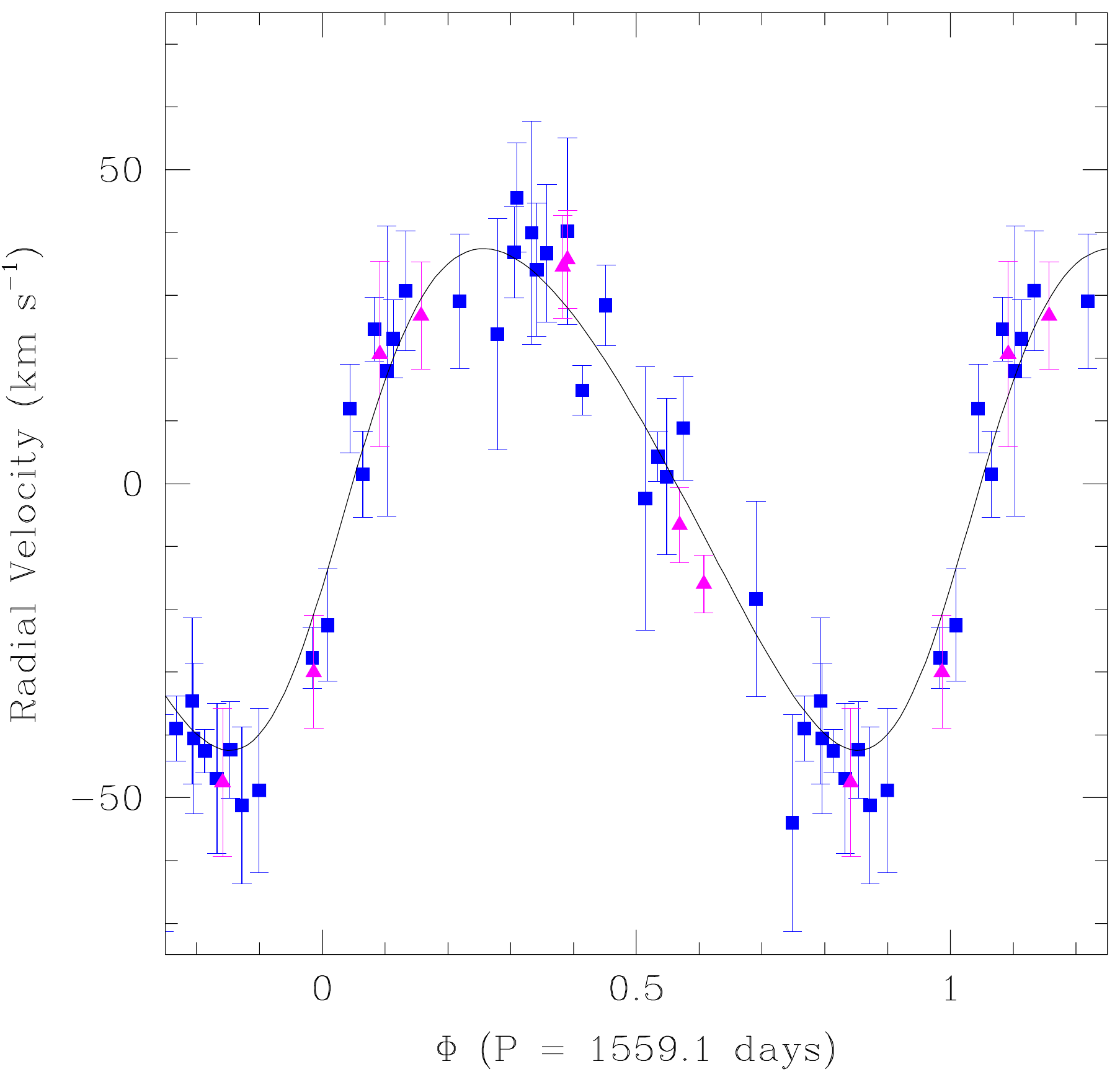}}
  \end{center}
  \caption{RVs of the He\,{\sc ii} emission lines of the WN6o star. The blue squares stand for the TIGRE data, while the magenta triangles indicate OHP data. The solid line corresponds to our orbital solution inferred from the highly-ionized nitrogen lines.\label{RVHeII}}
\end{figure}

We also examined the RVs of three He\,{\sc ii} emission lines (He\,{\sc ii} $\lambda\lambda$\,4542, 4686, 4859) measured on our TIGRE and OHP data. Again, we subtracted the mean values of the TIGRE RVs to shift all the lines to the same systemic velocity near 0.0\,km\,s$^{-1}$. For the OHP data, we again took the mean of the RVs measured during each observing run. The results are shown in Fig.\,\ref{RVHeII}, along with our orbital solution derived from the RVs of the highly-ionized nitrogen lines combined with the data of \citet{Dsilva}. As one can see on this figure, the RVs of the He\,{\sc ii} lines are fully consistent with the orbital solution inferred from the nitrogen lines, with only a marginally larger dispersion (rms of $7.7$\,km\,s$^{-1}$). 

Finally, we briefly addressed the question of short-term RV variations of the WN6o star. Based on a high-cadence monitoring of WR~138 over two weeks in August 2019, \citet{Dsilva} concluded that the N\,{\sc v} line RV varied with $\sigma = 5$\,km\,s$^{-1}$. This is very similar to what we found from our intensive OHP campaigns \citep[see also][]{Pal13}. To further quantify this, we applied our Fourier method to our complete time series of 136 RV data of highly-ionized nitrogen lines (i.e.\ not using the mean RVs of each OHP run, but considering instead all individual RVs). In this way, we checked that there is no signal in the RV time series at frequencies around the 2.3238\,day period advocated by \citet{Lamontagne} or at any other short period. We thus confirm the finding by \citet{Ann90} and \citet{Pal13} that there is no evidence for the presence of a close companion to the WN6o star.  

\subsection{The OB absorption lines: to move or not to move?}
\begin{figure}%[thb]
  \begin{center}
    \resizebox{7.5cm}{!}{\includegraphics{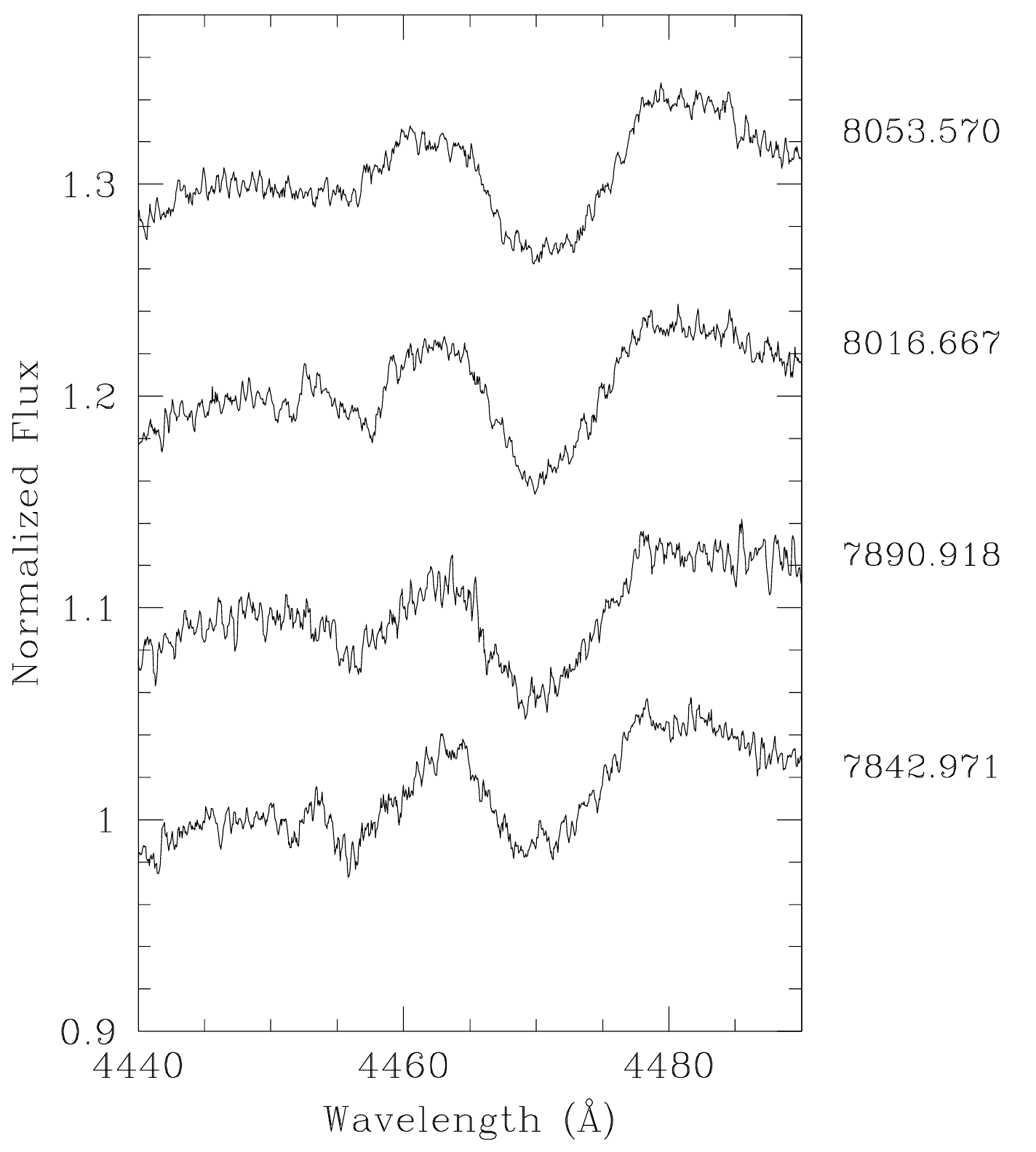}}
  \end{center}
  \caption{Example of the changing line profile of the He\,{\sc i} $\lambda$\,4471 absorption line. For clarity the different normalized spectra are vertically shifted. The data were taken with TIGRE in 2017 and the dates of the observations are indicated on the right in the format HJD-2450000.\label{montage}}
\end{figure}

As noted in previous studies \citep{Lamontagne,Ann90}, the RVs of the OB star are even more difficult to establish than those of the WN6o component. This is because the absorption lines are broad and fall on top of the WN6o emission lines. Moreover, they apparently undergo line profile variations (see Fig.\,\ref{montage}). The latter could be intrinsic to the OB star, or they could result from the blends with the WN6o emission lines which are themselves undergoing substantial line profile variations \citep{Lep99,Pal13}. As a first attempt to derive the RVs of the OB star, we measured the three strongest and least blended He\,{\sc i} absorption lines in the optical spectrum (He\,{\sc i} $\lambda\lambda$\,4471, 4922, 5876) by fitting Gaussian profiles. To evaluate RV$_{\rm OB}$ for a given epoch of observation, we took the mean of the RVs of those three lines from which we had previously subtracted the mean RV value of that line averaged over all TIGRE data. For the OHP data, we further averaged the values from the different nights of each observing run.

\begin{figure}%[thb]
  \begin{center}
    \resizebox{8.5cm}{!}{\includegraphics{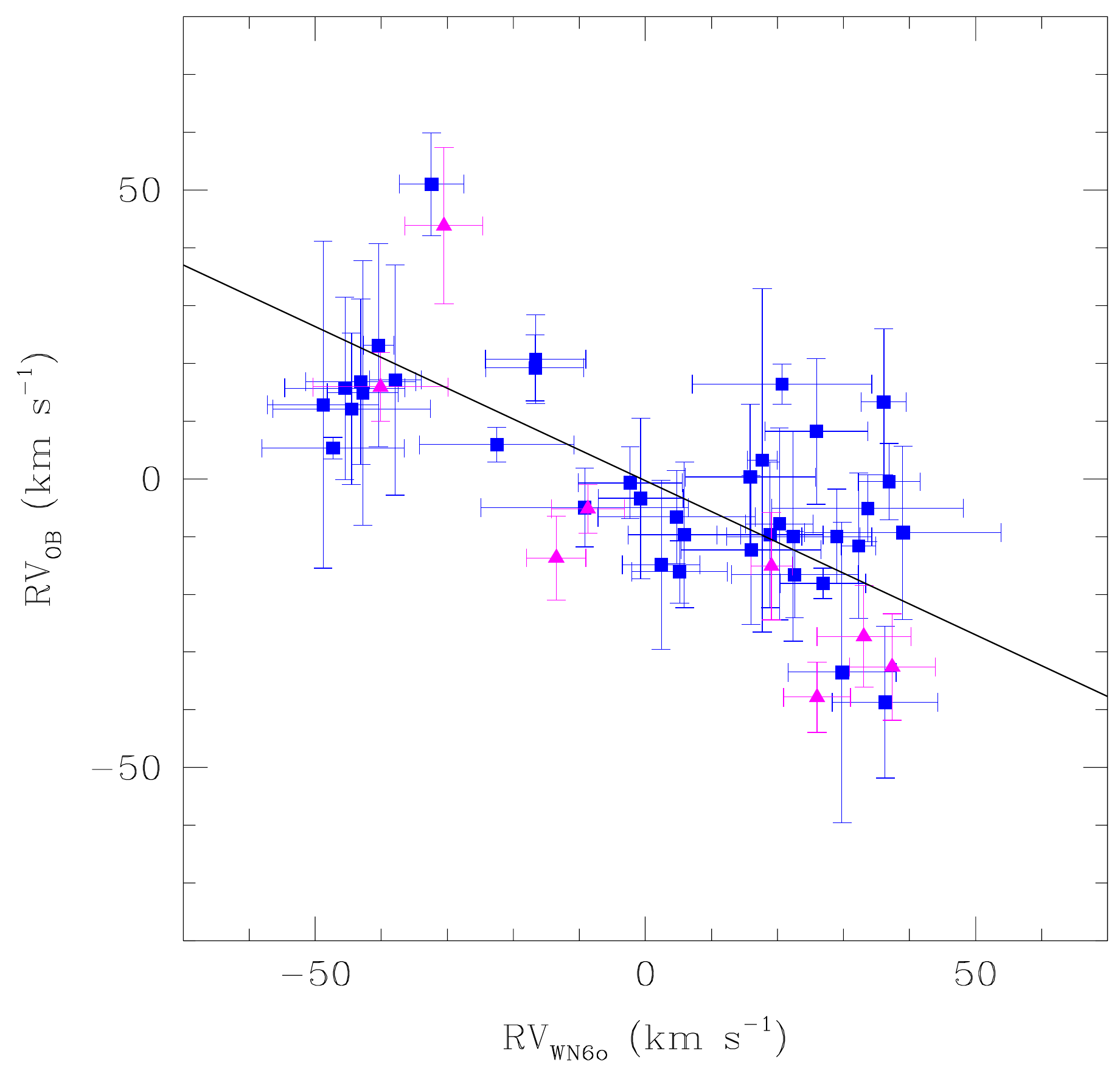}}
  \end{center}
  \caption{RV$_{\rm OB}$ as a function of RV$_{\rm WN}$ for our TIGRE (blue squares) and OHP (magenta triangles) data. The solid line corresponds to our best linear regression relation with $q = 0.53$.\label{massratio}}
\end{figure}

\begin{figure}
  \begin{center}
    \resizebox{8.5cm}{!}{\includegraphics{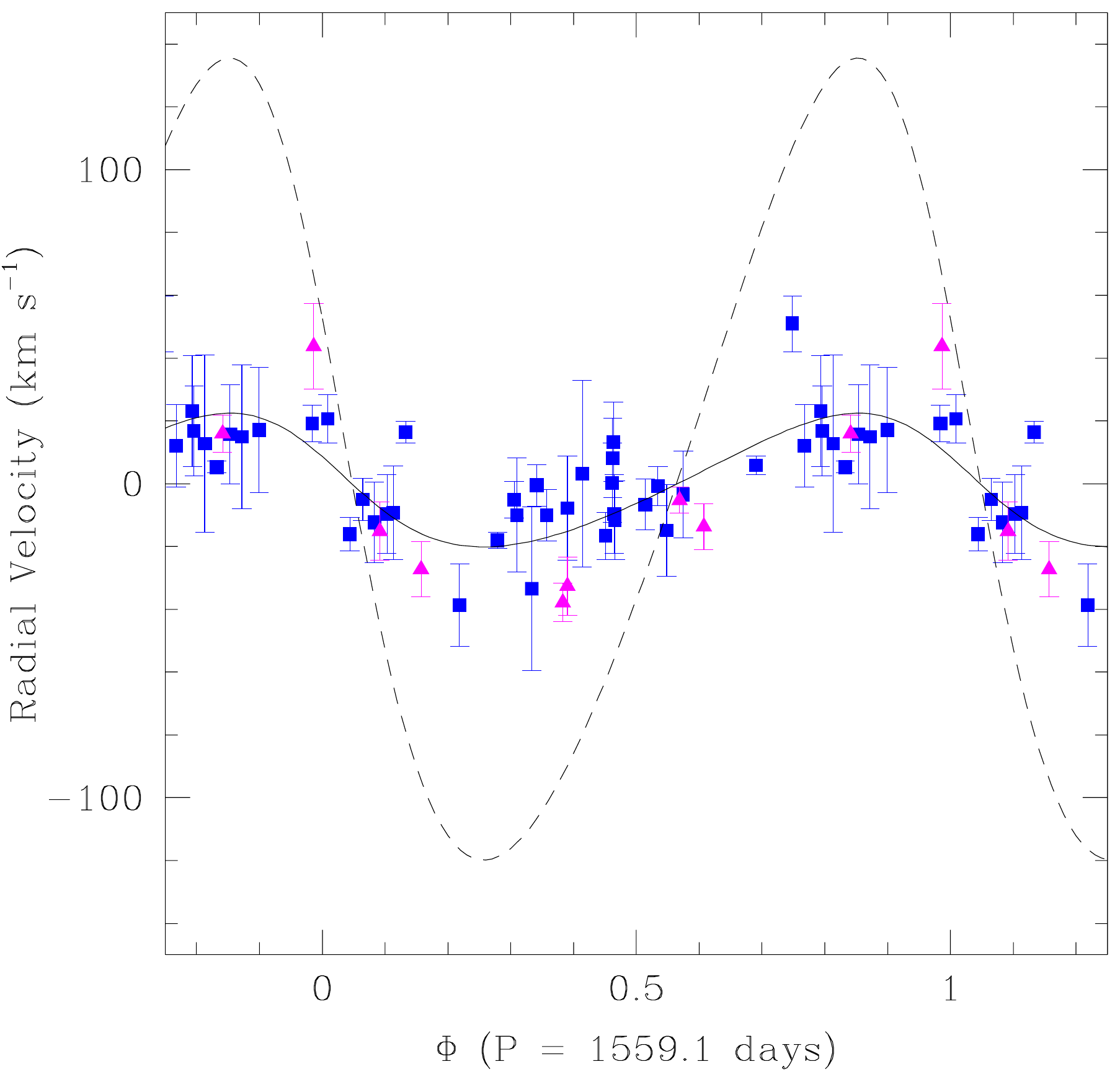}}
  \end{center}
  \caption{RV$_{\rm OB}$ along with the orbital solution corresponding to the reflex motion of the OB star around the WN6o star for $q = 0.53$ (solid line). TIGRE data are shown by blue squares, while OHP data are shown by magenta triangles. The dashed line corresponds to a value of $q = 3.20$ required to match the interferometric separation measured by \citet{Rich16} for an orbital inclination of $90^{\circ}$ and a distance of 1380\,pc (see text). \label{RVOB}}
\end{figure}

Overall, the values of RV$_{\rm OB}$ computed from the Gaussian fits are subject to $\sim 1.5$ times larger uncertainties than those of RV$_{\rm WN}$. Moreover, they show a significantly larger dispersion when folded with the orbital period. Figure\,\ref{massratio} illustrates RV$_{\rm OB}$ versus RV$_{\rm WN}$ along with a linear regression. The best regression yields a mass ratio of $q = \frac{m_{\rm WN6o}}{m_{\rm OB}} = 0.53 \pm 0.09$. The corresponding orbital solution is shown along with the RV$_{\rm OB}$ values in Fig.\,\ref{RVOB}.


Together with the mass function
$$f(m) = \frac{m_{\rm OB}\,\sin^3{i}}{(1 + q)^2} = (9.9 \pm 2.6)\,{\rm M}_{\odot}$$
from the SB1 orbital solution of the WN6o component, the value of $q$ results in minimum masses of
$m_{\rm WN6o}\,sin^3{i} = (12.3 \pm 4.2)$\,M$_{\odot}$ and $m_{\rm OB}\,sin^3{i} = (23.2 \pm 6.9)$\,M$_{\odot}$. Comparing the minimum mass of the OB star with the typical mass of O9.5\,V stars (23.4\,M$_{\odot}$) from \citet{Mar05}, we conclude that the orbital inclination must be rather close to $90^{\circ}$, and should be in any case larger than $\sim 62^{\circ}$.  

Since the blend of the OB absorption lines with the WN6o emission lines could bias our determination of the OB RVs, we made another attempt to establish these RVs by using our spectral disentangling code based on the shift and add algorithm described by \citet{Gon06}. This method allows in principle to reconstruct the individual mean spectrum of the components of a binary system and to simultaneously determine their RVs for each epoch of observation. In the present case, we fixed the RVs of the WN6o star to the values computed from the best SB1 orbital solution based on the N\,{\sc iv} and N\,{\sc v} emission lines. During the first 30 iterations, the RVs of the OB star were set to the RV$_{\rm OB}$ values described above. In the subsequent iterations, the RVs of the OB star were then allowed to vary. For each observation, we cross-correlated the residual spectrum, corresponding to the difference between the observed spectrum and the reconstructed WN6o spectrum shifted with the appropriate RV, with a TLUSTY synthetic spectrum\footnote{For this purpose, we used a synthetic spectrum with $T_{\rm eff} = 30\,000$\,K, $\log{g} = 4.0$ rotationally broadened to $v\,\sin{i} = 350$\,km\,s$^{-1}$.} \citep{Lan03}. This method was applied to data over three spectral ranges (4000 -- 4400\,\AA, 4440 -- 4580\,\AA, and 4800 -- 5000\,\AA) which display rather strong absorption lines in the synthetic spectrum. We restricted ourselves to the 24 TIGRE spectra with the highest S/N ratio. Yet, the results did not allow us to improve our estimates of the RV of the OB star. Indeed, for a given epoch of observation, the RVs of the OB star estimated from the different wavelength domains display a significant dispersion. This dispersion most likely arises from the short-term variability that affects the data and produces asymmetric line profiles (see Fig.\,\ref{montage}).

\begin{figure*}%[thb]
  \begin{minipage}{8.5cm}
  \begin{center}
    \resizebox{8.5cm}{!}{\includegraphics{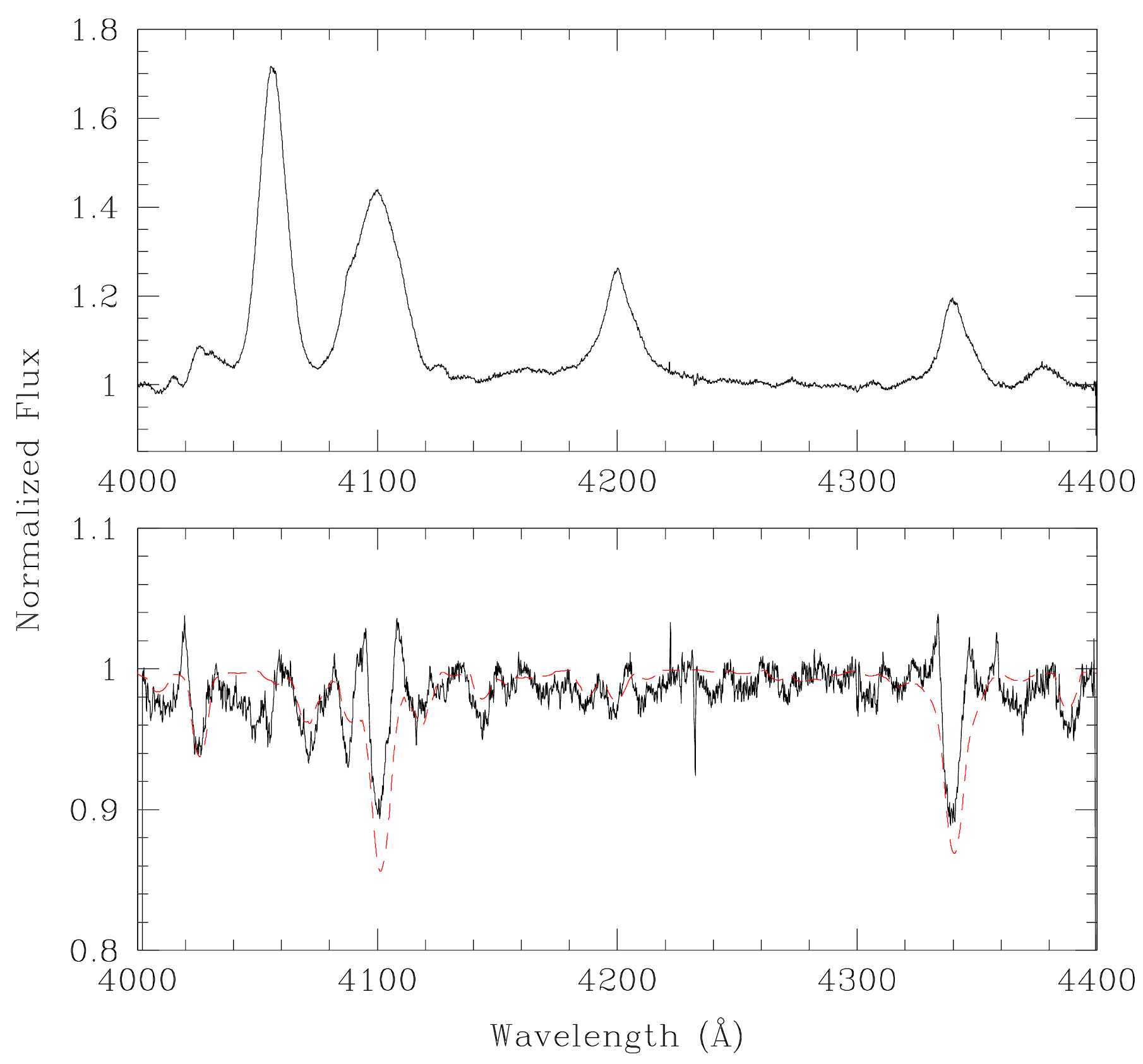}}
  \end{center}
  \end{minipage}
  \hfill
\begin{minipage}{8.5cm}
  \begin{center}
    \resizebox{8.0cm}{!}{\includegraphics{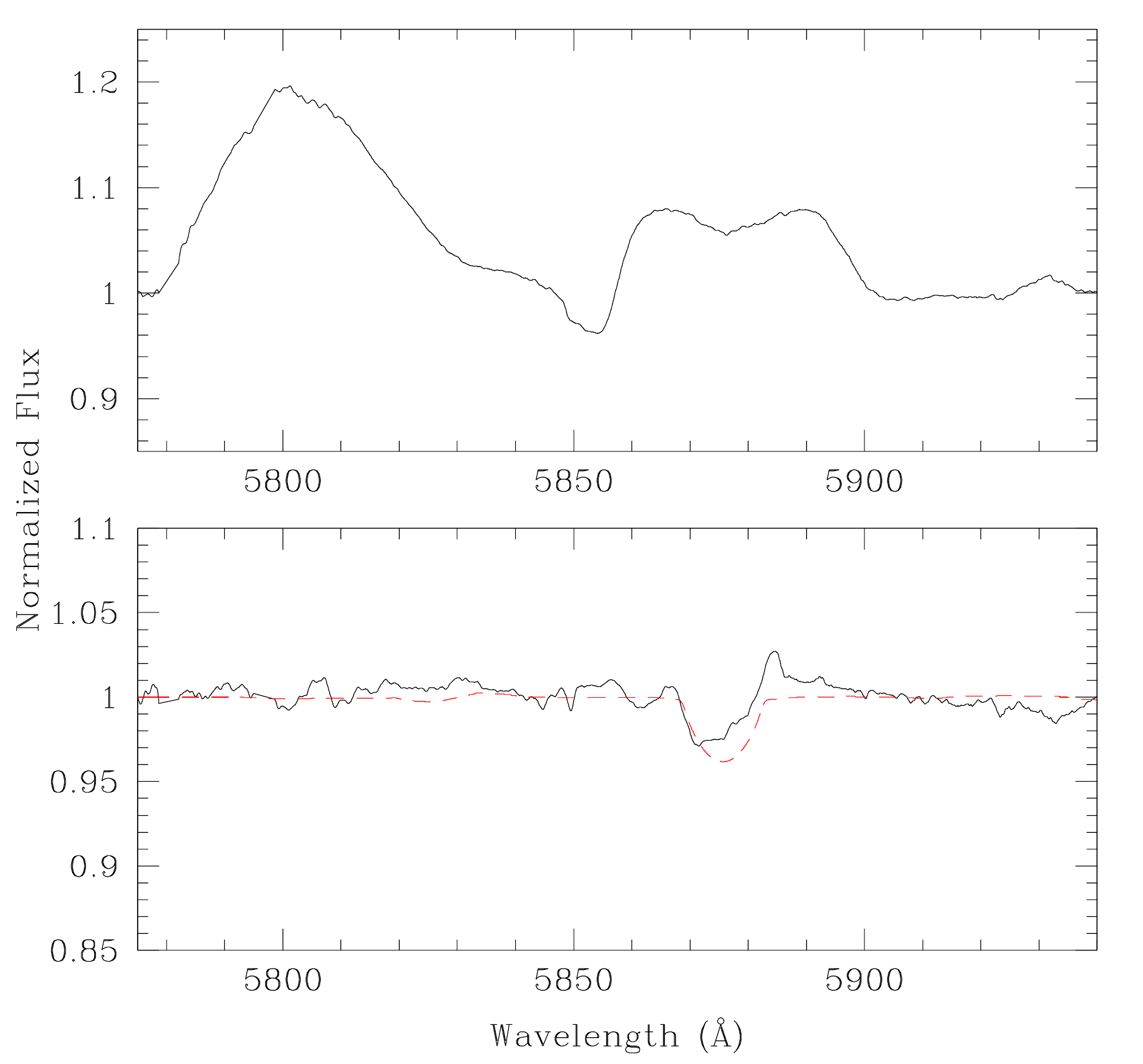}}
  \end{center}
\end{minipage}
\caption{Reconstructed spectra of the WN6o star (top panels) and the OB component (bottom panels) in the 4000 - 4400\,\AA\ spectral range (left) and in the 5775 - 5940\,\AA\ domain (right). These spectra were obtained via disentangling of the TIGRE data with the RVs set to the orbital solution of the WN6o star and $q = 0.53$. The red dashed line in the bottom panels illustrates a TLUSTY synthetic spectrum with $T_{\rm eff} = 32\,500$\,K, $\log{g} = 4.0$ and $v\,\sin{i} = 350$\,km\,s$^{-1}$, diluted by a factor 2. \label{disent}}
\end{figure*}

\begin{figure}
  \begin{center}
    \resizebox{8.0cm}{!}{\includegraphics{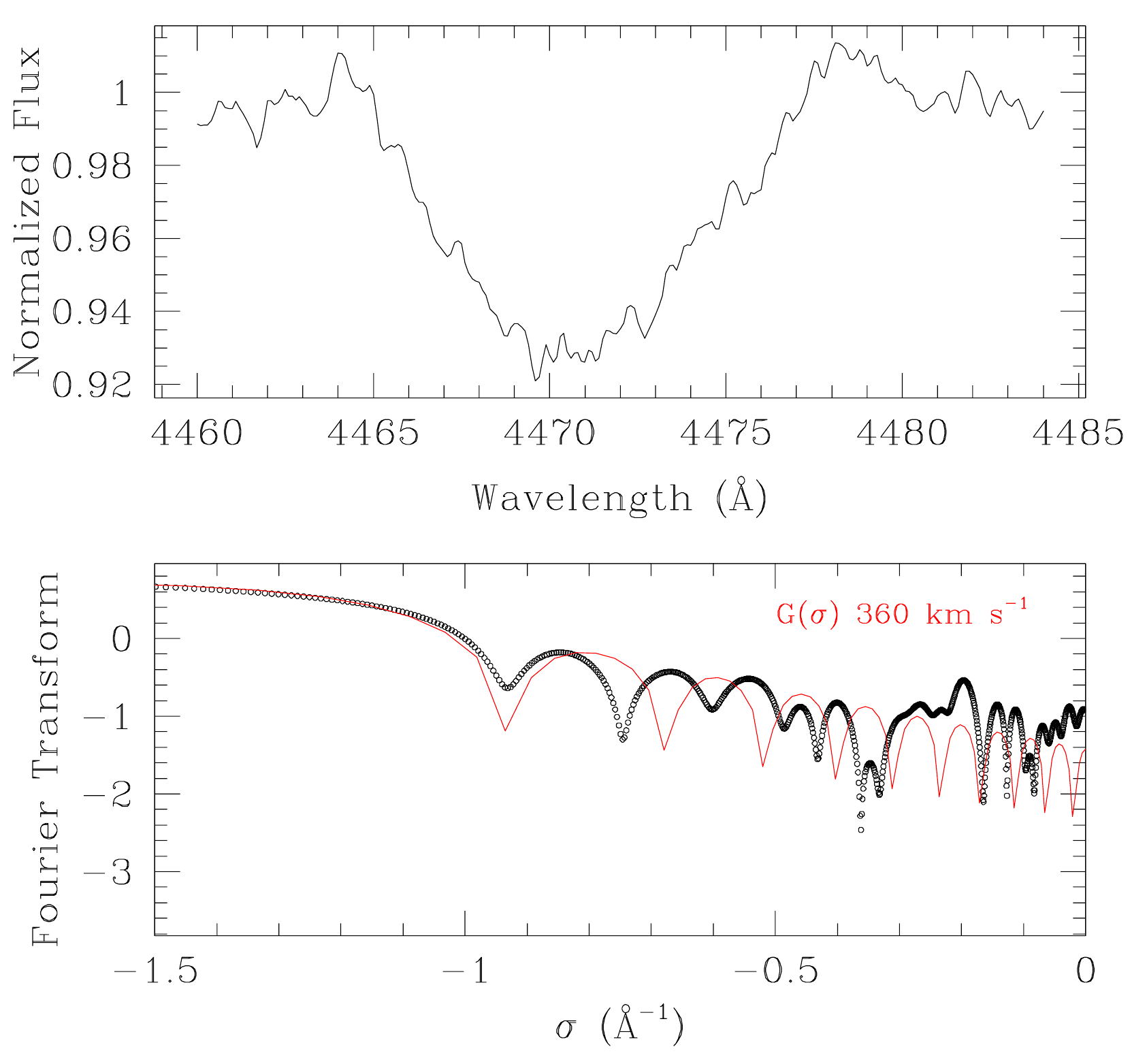}}
  \end{center}
  \caption{Top panel: He\,{\sc i} $\lambda$\,4471 absorption line of the OB star reconstructed via disentangling of the TIGRE data with the RVs set to the orbital solution of the WN6o star and $q = 0.53$. Bottom panel: Fourier transform of that line (black dots) and rotational profile (red curve) that best matches the position of the first zero of the Fourier transform.\label{vsini}}
\end{figure}

In view of the above-mentioned difficulties, we finally performed the spectral disentangling keeping the RVs of the WN6o star fixed to those of the best SB1 orbital solution and setting the RVs of the OB star to those corresponding to a mass ratio of $q = 0.53$. While the reconstructed spectrum of the OB star unveils absorption lines that were difficult to distinguish in the original individual observations (see left panel of Fig.\,\ref{disent}), the disentangling is certainly far from perfect. Indeed, we note that in some lines (e.g.\ He\,{\sc i} $\lambda\lambda$\,4471 and 5876, see right panel of Fig.\,\ref{disent}), the reconstructed spectrum of the WN6o star still displays some residual absorptions, implying that the strength of those lines is probably underestimated in the reconstructed spectrum of the OB star. Moreover, the variability of the WN6o spectrum due to wind inhomogeneities also pollutes the reconstructed spectrum of the OB star, notably in the He\,{\sc ii} $\lambda$\,4686 and the H\,{\sc i} Balmer lines which display strong emission artefacts.

Keeping those imperfections in mind, we have nevertheless attempted a rough spectral classification of the OB star based on the disentangled spectra. The ratio of the equivalent widths of the He\,{\sc i} $\lambda$\,4471 and He\,{\sc ii} $\lambda$\,4542 absorptions near $\sim 4$ yields a spectral type O9.5 \citep{Con71,Mat88}. Taking into account that the reconstructed OB spectrum might underestimate the strength of He\,{\sc i} $\lambda$\,4471, an O9.7 spectral type remains possible. An O9.5 spectral type is also in agreement with the fact that He\,{\sc ii} $\lambda$\,4200 is slightly weaker than He\,{\sc i} $\lambda$\,4144 \citep{Sot11}. In the same way, the ratio of the equivalent widths of the Si\,{\sc iv} $\lambda$\,4089 and He\,{\sc i} $\lambda$\,4144 absorptions suggests a V-III luminosity class \citep{Con71}. The Si\,{\sc iv} $\lambda$\,4089/He\,{\sc i} $\lambda$\,4026 criterion of \citet{Sot11} suggests a slightly more luminous III - II luminosity class. Comparing the strength of the He\,{\sc i} lines relative to the continuum with those of the same lines in synthetic TLUSTY spectra for $T_{\rm eff}$ equal to 30\,000 or 32\,500\,K, we conclude that the absorption lines in the spectrum of WR~138 are diluted by about a factor 2. Hence, the optical brightness of the OB star and the WN6o star should be nearly equal, in good agreement with the results from the model atmosphere spectral energy distribution (SED) fitting of \citet{Rich16}. We finally applied the Fourier transform method \citep{Sim07} to the He\,{\sc i} $\lambda\lambda$\,4471, 4922 line profiles to infer the value of the projected rotational velocity $v\,\sin{i}$ (see Fig.\,\ref{vsini}). In this way, we inferred values of $(360 \pm 10)$\,km\,s$^{-1}$ for the He\,{\sc i} $\lambda$\,4471 line, and $(340 \pm 10)$\,km\,s$^{-1}$ for He\,{\sc i} $\lambda$\,4922. Our results hence indicate a $v\,\sin{i} = (350 \pm 10)$\,km\,s$^{-1}$, in excellent agreement with \citet{Rich16}.     

\begin{figure}
  \begin{center}
    \resizebox{8.0cm}{!}{\includegraphics{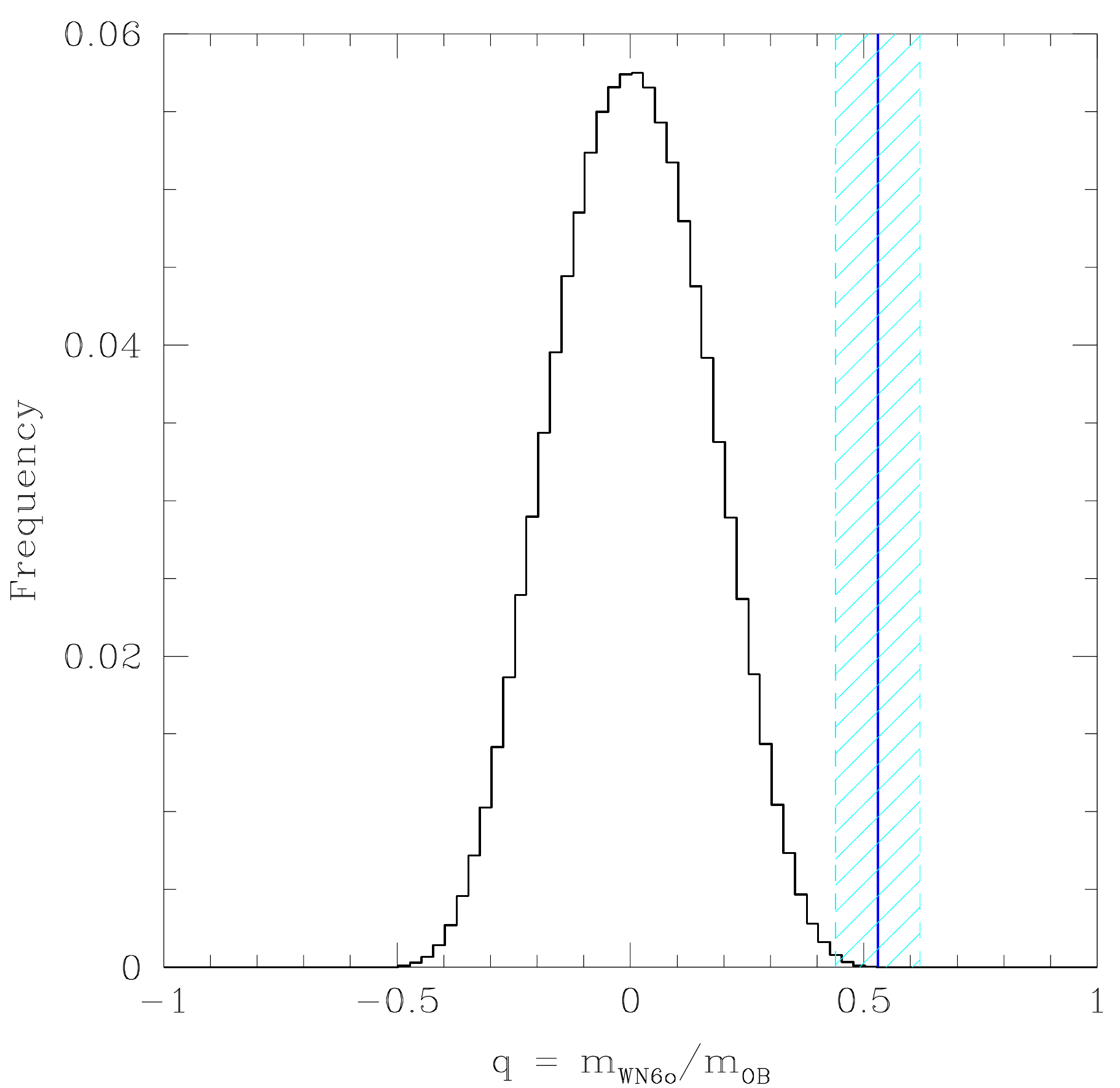}}
  \end{center}
  \caption{Histogram of the distribution of mass-ratios inferred from a million simulations of randomly shuffled pairs of RV$_{\rm OB}$ and RV$_{\rm WN}$ measurements. Our observational value is indicated by the solid blue line and its 1\,$\sigma$ uncertainty is shown by the hatched cyan area.\label{histoq}}
\end{figure}
Given the uncertainties on the RVs of the OB star, and the difficulties to reconcile the RV orbital solution and the interferometric results (see the forthcoming Sect.\,\ref{disc}), we further tested the possibility that the OB star might not be moving with the 4.3\,yr period, by setting the RVs of the OB star to zero in the disentangling process. The reconstructed spectra of the OB star looked quite similar to those obtained for $q = 0.53$ despite the radically different assumptions on the RVs. This situation prompted us to test the significance of the anti-correlation between RV$_{\rm OB}$ and RV$_{\rm WN}$ (Fig.\,\ref{massratio}). We first computed the Pearson correlation coefficient $r = -0.68$ and the Spearman rank-order correlation coefficient $r_S = -0.65$. Both values indicate a highly significant anti-correlation between RV$_{\rm OB}$ and RV$_{\rm WN}$. To go one step further, we performed a series of simulations where we re-shuffled the couples of RV$_{\rm OB}$ and RV$_{\rm WN}$ measurements, randomly assigning one of the observed RV$_{\rm OB}$ (and its estimated error) to an observed RV$_{\rm WN}$ (along with its error). For each simulated realization of our time series, we inferred the slope of a putative linear relation between RV$_{\rm OB}$ and RV$_{\rm WN}$, and thus the corresponding value of $q$. We repeated this procedure a million times and built a histogram of the frequency of the values of $q$. The result is illustrated in Fig.\,\ref{histoq}. We stress that negative values of $q$ represent positive correlations between RV$_{\rm OB}$ and RV$_{\rm WN}$ and do thus not reflect a genuine mass ratio. Because of the random shuffling, the expected mean value of the slope is zero (corresponding to the absence of any correlation). Comparing with our observed value of $q = 0.53 \pm 0.09$, we find that the re-shuffled time series yield a slope $\geq 0.53$ in only 0.0008\% of the simulations. A slope $\geq 0.44$ (i.e.\ $\geq 0.53 - 1\,\sigma$) is found for 0.08\% of the cases, and a slope $\geq 0.26$ (i.e.\ $\geq 0.53 - 3\,\sigma$) for 5.3\%. Hence, we conclude that the anti-correlation between RV$_{\rm OB}$ and RV$_{\rm WN}$ is indeed highly significant.
%We thus conclude that the spectra do not allow us to safely establish the existence of an orbital motion of the OB star with a period of 4.3\,yrs. 

\begin{figure*}%[thb]
  \begin{minipage}{8.5cm}
    \resizebox{8cm}{!}{\includegraphics{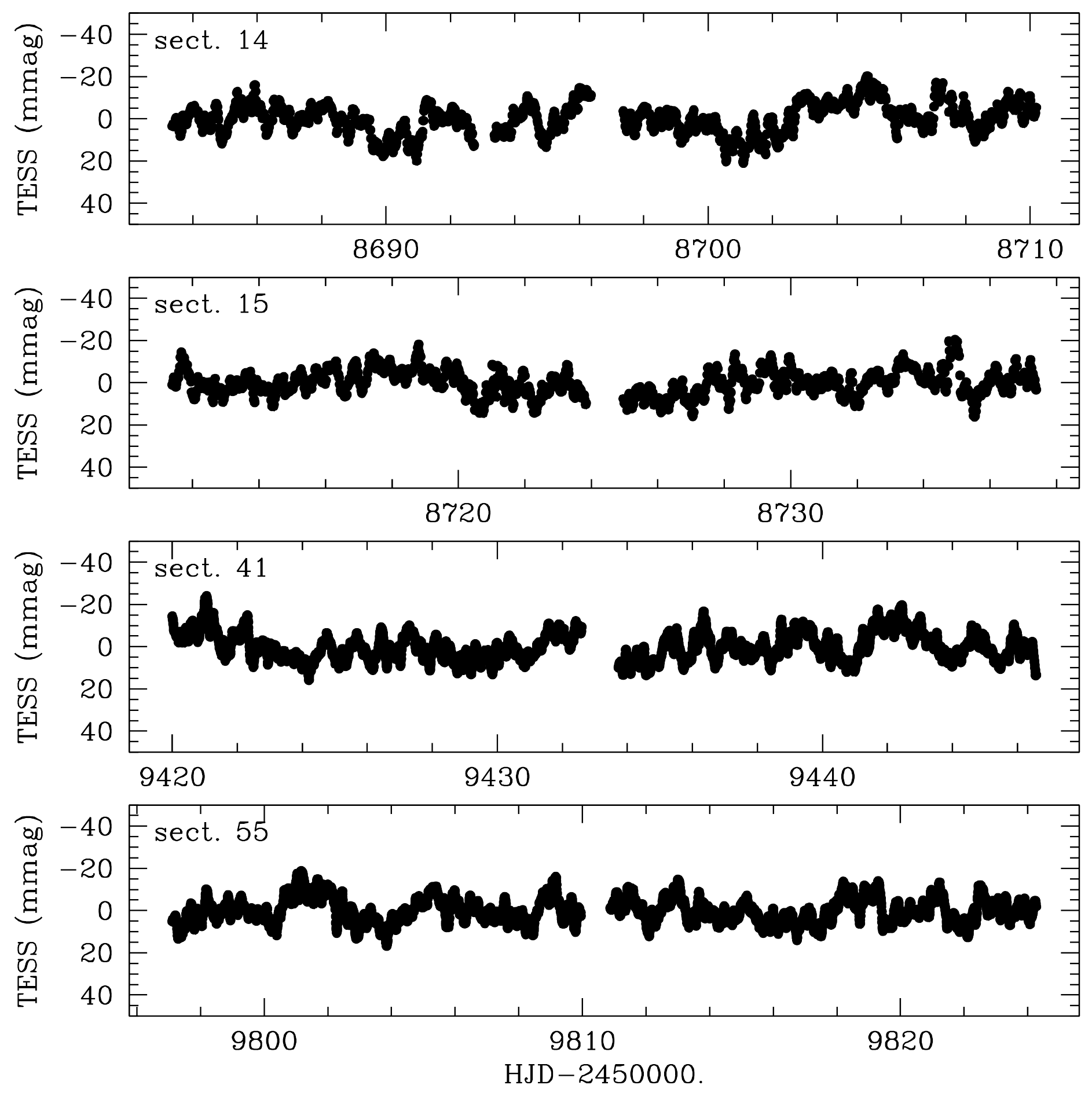}}
  \end{minipage}
  \hfill
  \begin{minipage}{8.5cm}
    \resizebox{8cm}{!}{\includegraphics{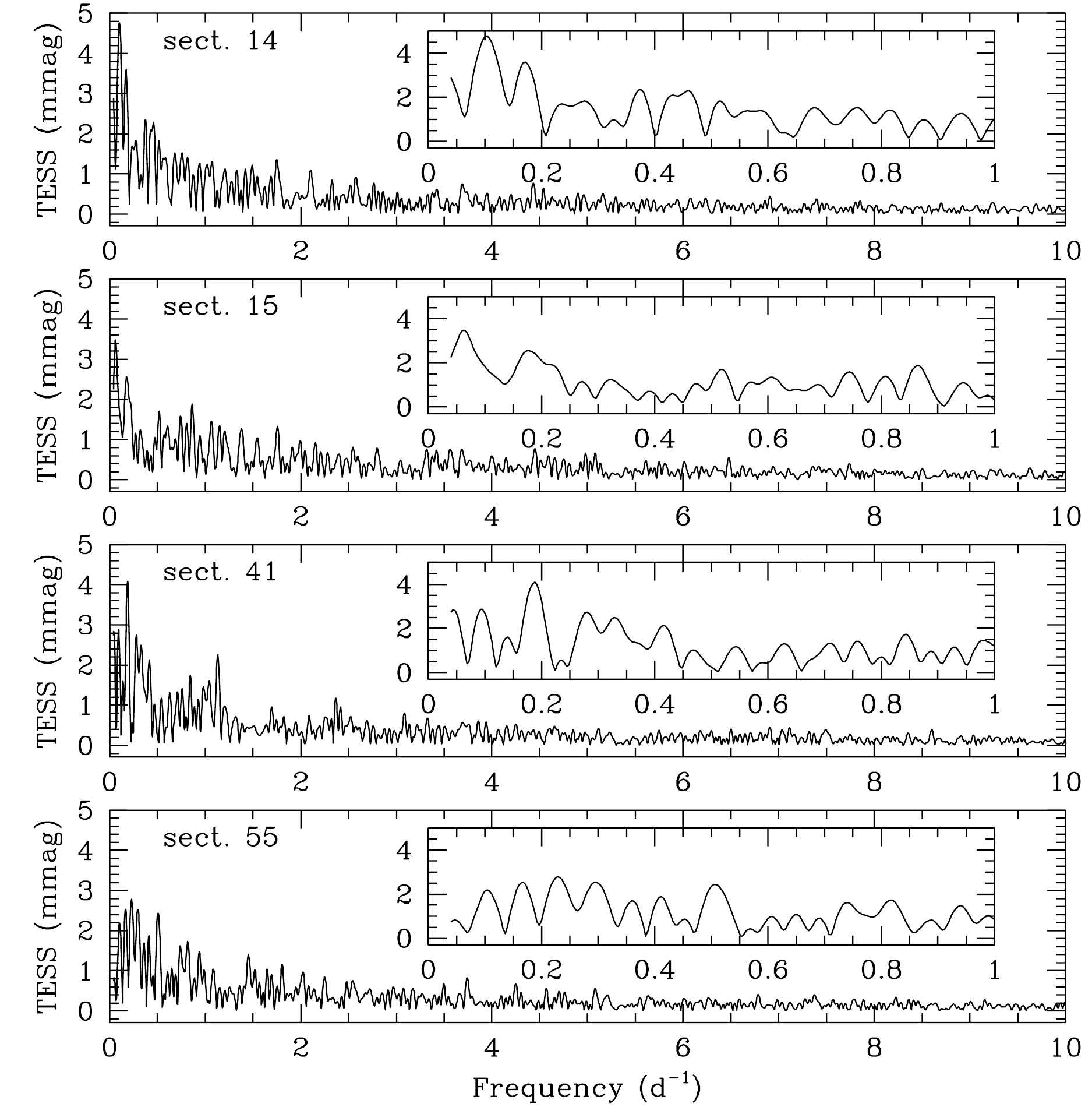}}
  \end{minipage}
  \caption{{Left: \it TESS} light curves of WR~138. Right: Fourier periodograms of the {\it TESS} photometry of WR~138. The inserts provide a zoom on the low-frequency part below 1\,d$^{-1}$.\label{TESS_lc}}
\end{figure*}

\section{Photometry \label{photom}}
\citet{MP}, \citet{Gap} and \citet{Ross} first detected low-level irregular photometric variability of WR~138. A photometric period of 11.6 days was reported by \citet{MS}. However, this result was based on a rather scarce data set and could be an artefact as sporadic periodicities are frequently found in ground-based photometric observations of WR stars \citep[e.g.][]{Gos94}. A first set of space-borne photometric measurements was obtained with the {\it Hipparcos} satellite. From these data, \cite{Mar98} reported WR~138 to display a stochastic variability.

The high-cadence {\it TESS} data allow us to reconsider this point. The left panel of Fig.\,\ref{TESS_lc} illustrates the light curve of WR~138 over the four sectors covered by {\it TESS}. From this figure, it becomes clear that the star displays variability at the $\sim 10$\,mmag level. No obvious periodic pattern is apparent in the light curves. To quantify this, we applied our Fourier period search method to the time series of each sector. The resulting Fourier periodograms are illustrated in the right panel of Fig.\,\ref{TESS_lc}.

A first important conclusion, is that there are no significant peaks that could hint at a periodic signal. Moreover, the content of the periodogram changes with time, indicating that we are seeing stochastic variability. We note also that these periodograms are essentially empty at frequencies above 4\,d$^{-1}$. Instead, the level increases towards lower frequencies, as one would expect for a variability dominated by red noise. This situation is not unexpected as red noise is commonly detected in the analyses of space-borne photometry of massive stars, including Wolf-Rayet stars \citep[][and references therein]{Naz21}.

In the case of WR~138, roughly half of the light collected in the {\it TESS} waveband (i.e.\ between 6000\,\AA\ and 1\,$\mu$m) should come from the WN6o star and about half from the OB star \citep{Rich16}. The absence of significant stable periodicities indicates that the OB companion is probably not a $\beta$~Cep-like pulsator. Therefore, the line profile variability that affects the broad He\,{\sc i} absorption lines is probably not due to stable pulsations, but rather to stochastic variations of the OB star and wind variability in the spectrum of the WN6o star.

\section{Discussion \label{disc}}
\subsection{Comparison with interferometric data}
As pointed out in Sect.\,\ref{intro}, \citet{Rich16} resolved WR~138 with the {\it CHARA} interferometer into two objects separated by 12.4\,mas. The $H$-band brightness ratio of the two sources inferred from the {\it CHARA} data was in agreement with the two objects corresponding to the WN6o and OB stars. The date of these interferometric observations corresponds to orbital phase 0.43 in our best SB1 orbital solution of the WN6o star. At that phase, the orbital separation amounts to $r = 1.15\,a$ where $a$ is the semi-major axis of the orbit. Likewise, the position angle $\psi$ (defined as $0^{\circ}$ at inferior conjunction of the WN6o star) was $141^{\circ}$. The projection of the separation of the two stars on the sky then amounts to
$$\sqrt{(r\,\cos{\psi}\,\cos{i})^2 + (r\,\sin{\psi})^2} \cdot$$
If we adopt $q = 0.53$, our best orbital solution leads to $a\,\sin{i} = (1859 \pm 120)$\,R$_{\odot}$, and thus $r\,\sin{i} = (2138 \pm 149)$\,R$_{\odot}$.

The most recent Gaia parallax of WR~138 is equal to $\varpi = 0.449 \pm 0.022$\,mas \citep{EDR3}, corresponding to a distance of $2.13 \pm 0.09$\,kpc \citep{Bai21} or $2.2 \pm 0.1$\,kpc \citep{Cro23} depending on the adopted priors. Adopting a distance of 2.15\,kpc, the projected separation at the time of the {\it CHARA} observations was thus seen under an angle of 
$$\frac{(4.6 \pm 0.7)\,{\rm mas}}{\sin{i}}\,\sqrt{(\cos{\psi}\,\cos{i})^2 + (\sin{\psi})^2} \cdot $$
Assuming this quantity to be equal to 12.4\,mas, and solving for $i$ yields an inclination close to $20.9^{\circ}$ which would result in unrealistically high masses of over 500\,M$_{\odot}$ for the OB star, and 270\,M$_{\odot}$ for the WN6o star. From their SED fitting, \citet{Rich16} advocate a shorter distance of 1380\,pc. If we adopt this value, we must replace the 4.6\,mas in the relation above by 7.2\,mas. This then leads to an inclination of $32^{\circ}$ which would still imply masses of 156 and 83\,M$_{\odot}$, which are still clearly unrealistic. An alternative way to illustrate this problem is to apply Kepler's third law assuming that $P_{\rm orb} = 1559$\,days holds for the two objects separated by 12.4\,mas, and that this angular separation corresponds to $a$ (i.e.\ neglecting projection effects and the small eccentricity). Adopting a distance of 2.15\,kpc would imply a total mass of the binary of 980\,M$_{\odot}$, whilst the shorter distance of 1380\,pc would still lead to a total mass of 260\,M$_{\odot}$. Yet another way to illustrate this problem is provided in Fig.\,\ref{RVOB}. The dashed line in this figure illustrates the RV curve of the OB star one would expect assuming the interferometric separation corresponds to the on-sky projection of the orbital separation of the system with a 4.3\,yr orbit. In Fig.\,\ref{RVOB}, we have assumed an orbital inclination of $90^{\circ}$ and a distance of 1380\,pc \citep{Rich16}. This would imply a mass ratio of $q = 3.20$ which is clearly at odds with the values of RV$_{\rm OB}$, and would once again lead to unrealistic masses of 174\,M$_{\odot}$ for the OB star and 558\,M$_{\odot}$ for the WN6o star. Adopting instead the Gaia distance would lead to an even higher value of $q$, making the situation even worse.  

Therefore, we conclude that it is currently impossible to reconcile the results of the {\it CHARA} observation of \citet{Rich16} with our spectroscopic RV solution. Either the interferometric solution is somehow biased, or we are forced to conclude that the object resolved with {\it CHARA} is not the star responsible for the orbital motion of the WN6o star with a period of 1559\,days, but a third component that does not undergo significant RV variations over our observing campaign. At this stage, we do not know whether this third component is an early-type star or not, but in case it is, it will likely contribute to the He\,{\sc i} lines we have measured. This could explain some of the difficulties encountered in establishing the orbital motion of the OB component of the 4.3\,yr binary.

We can wonder whether a physical triple configuration would be stable. Assuming that the inner binary consists of a WN6o star of $\sim 12.5$\,M$_{\odot}$ orbited every 1559\,days by a 23.5\,M$_{\odot}$ O9.5\,V star and that this system is itself orbited by another O9.5\,V star (the object detected with {\it CHARA}) of identical mass, we can estimate a total mass of the triple system of $\sim 60$\,M$_{\odot}$. Kepler's third law then implies minimum orbital periods for the outer orbit of 6467\,days for a distance of 2.15\,kpc and 3326\,days for a distance of 1380\,pc. Whilst these are minimum values of the orbital period (due to the projection effect), they lead to a very small ratio between the orbital period of the outer and inner orbit, which would be inconsistent with a stable hierarchical triple system \citep{Tok04}. 

The luminosities and general properties of the components of WR~138 inferred by \citet{Rich16} were obtained assuming two stars (the WN6o star and its OB companion) at a distance of 1.38\,kpc. If we are dealing instead with three stars (the WN6o + OB binary system and a more distant third object), then scaling the luminosities found by \citet{Rich16} to the Gaia distance of 2.15\,kpc, the total luminosity of the binary system becomes $5.4\,10^5$\,L$_{\odot}$. Adopting the effective temperature (31\,000\,K) obtained by \citet{Rich16} and a mass close to 24\,M$_{\odot}$ for the OB star, the evolutionary models of \citet{Eks12} yield $\log{L/L_{\odot}} \sim 5.2$ for the OB star, suggesting a giant luminosity class \citep{Mar05}. Our best-fit mass-ratio then implies a mass of the WN6o star of $(12.7 \pm 1.1)$\,M$_{\odot}$. Using the model-based mass-luminosity relation for Wolf-Rayet stars proposed by \citet{Sch92}, we estimate $\log{L/L_{\odot}}$ between 5.31 and 5.56 for the WN6o star. This results in a combined luminosity of the WN6o + OB binary between $3.6\,10^5$\,L$_{\odot}$ and $5.2\,10^5$\,L$_{\odot}$, in reasonable agreement with the above estimate. 

\subsection{Evolutionary state}
At this stage, we need to return to the question whether the properties of WR~138 can be explained by binary interactions. Indeed, \citet{Rich16} suggested that WR~138 might have evolved through a RLOF episode with mass and angular momentum transfer despite its wide current orbital separation. For moderately wide binary systems with orbital periods in the range between 10 and 1000 days, so-called case B RLOF occurs while the mass donor is burning hydrogen in a shell (i.e.\ prior to core He burning). This scenario could indeed apply to WR~138. Three properties are worth considering in this context: the rapid rotation of the OB star, the orbital period and orbital eccentricity. 

\citet{Sha17} discussed the projected rotational velocities of the O-type companions in eleven WR + O binary systems with orbital periods between 4.2 and 78.5\,days. Accounting for the known inclinations of these systems, and assuming the rotation and orbital axes to be aligned, they inferred equatorial velocities that clearly exceed those expected for synchronous rotation and revolution. For the He\,{\sc i} lines, \citet{Sha17} derive an average $v_{\rm eq, He\,I}$ of 348\,km\,s$^{-1}$, whereas the He\,{\sc ii} lines yield a lower average of $v_{\rm eq, He\,II} = 173$\,km\,s$^{-1}$. \citet{Sha17} attribute this difference to the effect of gravity darkening in stars that are heavily flattened due to their rapid rotation. Unfortunately, the He\,{\sc ii} lines of the OB star in WR~138 are too weak to infer $v\,\sin{i}$ from the disentangled spectrum. Still, WR~138 shares many of the properties of the objects discussed by \citet{Sha17}. Indeed, considering the OB star to be the object that orbits the WN6o star every 4.3\,yrs, we find that its rotational velocity is clearly faster than synchronous. Actually, with its significantly longer orbital period, WR~138 is clearly a more extreme case than the objects in the sample of \citet{Sha17}. This is true as far as the orbital separation and thus the size of the Roche lobe are concerned, and also in terms of the subsequent evolution of the mass and angular momentum gainer.
Concerning the orbital separation, \citet{Rich16} argued that the orbital period of WR~138 is close to the upper limit on the binary period for an O-type star to undergo spin-up in a RLOF interaction \citep{San12}. Yet, depending on the amount of mass lost in the process, the initial orbital period could well have been shorter than the current value.

If the case B mass transfer is quasi conservative, the mass gainer would be spun-up to near critical velocity \citep{Pac81}. Yet, the rotational velocities inferred by \citet{Sha17} and also that of the OB star in WR~138 are significantly below the critical rotation rate. In close systems with short orbital periods, this situation could stem from tidal interactions that slow down the O star's rotation. For longer period systems, especially for a long-period system such as WR~138, tidal interactions are much less efficient and thus cannot explain this result. \citet{Van18} accordingly suggested that differential rotation due to accretion of mass onto the O star could induce a Spruit-Taylor dynamo effect that would lead to a spin-down of the O star, resulting in a significantly subcritical rotational velocity at the end of the RLOF episode. In the \citet{Van18} scenario, one would thus expect the OB star to be magnetic. In this context, it is worth mentioning that \citet{dlC14} reported a weak spectropolarimetric signal in WR~138 that they interpreted as a marginal detection of an 80\,G magnetic field in the wind of the WN6o star. Yet, it is unclear whether this spectropolarimetric signal, which extends over $\sim 2000$\,km\,s$^{-1}$ in velocity space, could instead come from the OB component.  

Finally, while the eccentricity of the orbit of the WN6o star is relatively small, it is actually not zero. Quite frequently, it is assumed that a mass transfer episode should lead to rapid circularization of the orbit. However, \citet{DeG91} found that the properties of WR~140 (WC7 + O4-5V, $e \simeq 0.8$, P$_{\rm orb} = 7.9$\,yr) were best explained by a case B mass exchange modulated by the large eccentricity. He argued that tidal interactions were too slow to circularize the orbit and that the initial eccentricity of the binary was likely larger than the current value. \citet{Sep07,Sep09}, and \citet{Dos16} investigated the secular variations of the orbital eccentricity and semi-major axis during mass-transfer in eccentric binaries. They showed that these quantities can decrease or increase, depending on the initial values of the mass-ratio and the eccentricity. Based on these results, we can assume that a similar scenario might apply to the case of WR~138: it is likely that the initial eccentricity was considerably higher than the current value, but tidal interactions were not efficient enough to circularize the orbit. An alternative scenario that could apply if WR~138 eventually turned out to be a physical (hierarchical) triple system is Kozai-Lidov cycles. Indeed, in hierarchical triples, the Kozai-Lidov mechanism can lead to oscillations of the orbital eccentricity of the inner binary system \citep{Naoz}. Such oscillations can enhance tidal interactions and mass-exchange, thus influencing the evolution of the inner binary \citep{Too16}. However, at this stage, this is highly speculative as further observations are required to firmly establish or discard the triple star scenario. 

\section{Conclusions \label{conclusion}}
Our results indicate that the WN6o star in WR~138 is orbited by a relatively massive companion in a 4.3\,yr orbit with an eccentricity of 0.16. The rapidly rotating OB star that leaves its signatures in the combined spectrum offers the most plausible candidate for being this companion though its RVs are more difficult to measure. We have shown that the OB star RVs are anticorrelated with those of the WN6o star. It appears clearly that the companion of the WN6o star on the 4.3\,yr orbit cannot be the object that was resolved in interferometry with the {\it CHARA} array \citep{Rich16}. This opens up the possibility of a triple system and calls for additional interferometric observations.

\section*{Acknowledgements}
This work is based on spectra collected with the TIGRE telescope (La Luz, Mexico) and at the Observatoire de Haute Provence (France). TIGRE is a collaboration of Hamburger Sternwarte, the Universities of Hamburg, Guanajuato, and Li\`ege. This publication also uses data obtained with the {\it TESS} mission, whose funding is provided by the NASA Explorer Program. We thank the Belgian Federal Science Policy Office (BELSPO) for their support in the framework of the PRODEX Programme. The ADS, CDS and SIMBAD databases were used in this work.

\end{document}